# Science cited in policy documents: Evidence from the Overton database

Zhichao Fang[1*], Jonathan Dudek[1], Ed Noyons[1], Rodrigo Costas[1,2]

* Corresponding author

Zhichao Fang (ORCID: 0000-0002-3802-2227)
[1] Centre for Science and Technology Studies (CWTS), Leiden University, Leiden, The Netherlands.
E-mail: z.fang@cwts.leidenuniv.nl

Jonathan Dudek (ORCID: 0000-0003-2031-4616)
1 Centre for Science and Technology Studies (CWTS), Leiden University, Leiden, The Netherlands.
E-mail: j.dudek@cwts.leidenuniv.nl

Ed Noyons (ORCID: 0000-0002-3103-2991)
[1] Centre for Science and Technology Studies (CWTS), Leiden University, Leiden, The Netherlands.
E-mail: noyons@cwts.leidenuniv.nl

Rodrigo Costas (ORCID: 0000-0002-7465-6462)
[1] Centre for Science and Technology Studies (CWTS), Leiden University, Leiden, The Netherlands.
[2] DST-NRF Centre of Excellence in Scientometrics and Science, Technology and Innovation Policy, Stellenbosch University, Stellenbosch, South Africa.
E-mail: rcostas@cwts.leidenuniv.nl

**Abstract**

To reflect the extent to which science is cited in policy documents, this paper explores the presence of policy document citations for over 18 million Web of Science-indexed publications published between 2010 and 2019. Enabled by the policy document citation data provided by Overton, a searchable index of policy documents worldwide, the results show that there are 3.9% of publications in the dataset cited at least once by policy documents. Policy document citations present a citation delay towards newly published publications and show a stronger predominance to the document types of review and article. Based on the Overton database, publications in the field of Social Sciences and Humanities have the highest relative presence in policy document citations, followed by Life and Earth Sciences and Biomedical and Health Sciences. Our findings shed light not only on the impact of scientific knowledge on the policy-making process, but also on the particular focus of policy documents indexed by Overton on specific research areas.

**Introduction**

Since Altmetric.com incorporated policy document citations as one of the indicators for measuring the broader impact of research outputs, there have been a range of studies investigating the science cited in policy documents by relying on the data provided by Altmetric.com (Bornmann et al., 2016; Fang et al., 2020; Haunschild & Bornmann, 2017; Yu et al., 2020). In addition to Altmetric.com, Overton[1], a new database of policy documents, might open a new window for observing policy document citations to scientific publications. Overton claims to be the world's largest collection of policy documents, covering over a thousand sources worldwide (Adie, 2020b). The scope of policy documents defined by Overton is relatively broad because it tracks diverse sources, including governments, think tanks, NGOs, and IGOs (Wilson, 2020).

To date, Overton has not been explored at scale regarding the extent to which scientific publications are cited by policy documents, which is essential for better understanding the potential impact that scientific outcomes can have on the policy-making process. Thus, using the policy document data provided by Overton, this study fills this research gap by presenting an overview of the presence of policy document citations for scientific publications across different publication years, document types, and subject fields.

**Data and methods**

*Dataset*

Up to June 2020, the Overton database has indexed over 2.73 million policy documents on a global level, and about 14.5% of them have at least one cited scientific publication detected. Since policy documents in the Overton database are only sparsely distributed before 2009 (Adie, 2020a), we selected a total of 18,343,140 Web of Science (WoS) publications with DOIs published between 2010 and 2019 as the main dataset. To investigate whether or not they are cited by policy documents, these WoS publications were matched with the reference lists of policy documents tracked by Overton up to June 2020 based on their DOIs. Bibliometric information of the publications in our dataset were harvested from the CWTS in-house database, which is an enhanced version of the Science Citation Index (SCI), Social Science Citation Index (SSCI), Arts & Humanities Citation Index (A&HCI) and Conference Proceedings Citation Index (CPCI), primarily including their publication year and document type. The overall distribution of publications across publication years and document types can be found in Table A1 and Table A2 in the appendix, respectively.

*The CWTS publication-level classification*

To examine how the presence of policy document citations varies across subject fields, we employed the CWTS classification system (also known as the Leiden Ranking classification) to assign publications with their explicit subject fields. The CWTS classification is a publication-level classification system developed by Waltman and Van Eck (2012). As shown in Figure 1, by making use of the base map of the CWTS classification in its 2020 version, publications in our dataset are clustered into 4,013 micro-level fields (network nodes in Figure 1) with similar research topics, and these micro-level fields are further grouped into five main subject fields of science obtained algorithmically, including *Social Sciences and Humanities* (SSH), *Biomedical and Health Sciences* (BHS), *Physical Sciences and Engineering* (PSE), *Life and Earth Sciences* (LES), and *Mathematics and Computer Science* (MCS).[2] Since only certain citable document types (including Article, Proceeding Paper, Review, and Letter) are included in the CWTS classification, there are 15,926,452 publications (accounting for 86.8%) in our dataset assigned to a subject field. This set

---

[1] See more introduction to Overton at: https://www.overton.io/ (Accessed October 6, 2020)
[2] See more details about the CWTS classification at: https://www.leidenranking.com/information/fields (Accessed October 6, 2020)

of publications was drawn as a subsample to analyze the subject field variations of policy document citations. Table A3 in the appendix provides the distribution of publications across the five subject fields.

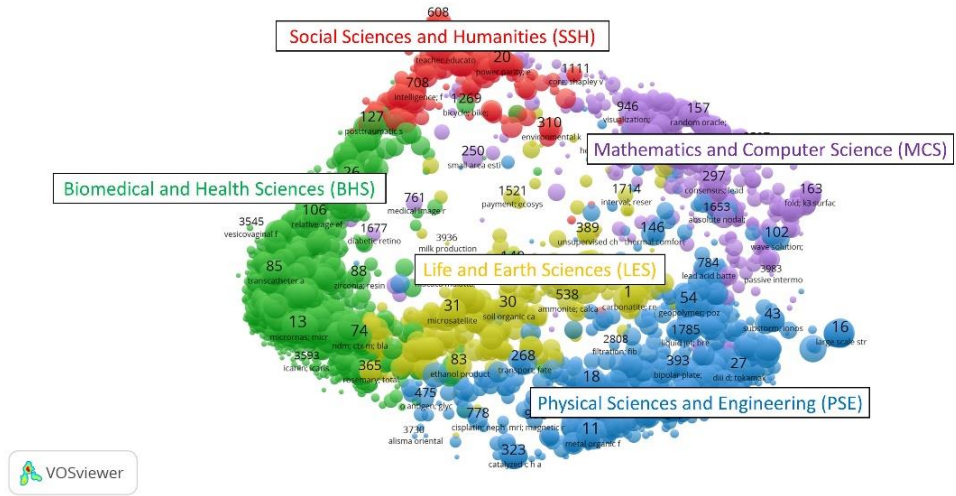

Figure 1. The layout of the five main subject fields of science in the CWTS classification system

*Indicators and analytical approaches*

To measure the presence of policy document citations, we used the three indicators proposed by Haustein et al. (2015): *Coverage*, *Density* and *Intensity*. For a specific set of publications, Coverage (C) refers to the percentage of publications with at least one policy document citation. The higher the value of Coverage, the higher the proportion of publications that received policy document citations. Density (D) is the average number of policy document citations received by all publications (including both publications with policy document citations and those without any policy document citations). The higher the value of Density, the more policy document citations received on average by the whole set of publications. Intensity (I) is the average number of policy document citations received by those publications with at least one policy document citation. The higher the value of Intensity, the more policy document citations received on average by those publications that attracted any policy attention.

VOSviewer was utilized as the visualization tool, and its embedded function of overlay visualization was employed to emphasize which micro-level fields caught comparatively more attention from the policy documents indexed by Overton.

**Results**

For the entire dataset of 18,343,140 publications in our study, the Coverage, Density and Intensity of policy document citations recorded by Overton are 3.9%, 0.09 and 2.32, respectively. In other words, there are 3.9% of publications received at least one policy document citation, and on average, publications in the dataset have been cited 0.09 times. Those publications that have been cited by policy documents at least once received 2.32 policy document citations on average.

Figure 2 shows the presence of policy document citations over the publication years of scientific publications from the perspectives of Coverage, Density and Intensity. The analysis shows an obvious downtrend over time. The longer the period of time that publications have been published, the higher the presence of policy document citations, whereas more recent publications have received sparser policy document citations. This indicates that older publications receive more citations in policy documents, in line with other types of *slow pace* accumulation metrics such as scholarly citations, Wikipedia citations or Peer review mentions (Fang & Costas, 2020). Take Coverage as an example, the share of publications cited by policy documents ranges from 0.78% for publications published in 2019 to 6.47% for those published in 2010. Thus, policy documents could be described as exhibiting a citation delay that is comparable to what can be observed with scientific publications as well: Usually, it takes a relatively long time for newly published publications to accumulate citations in other scientific publications.

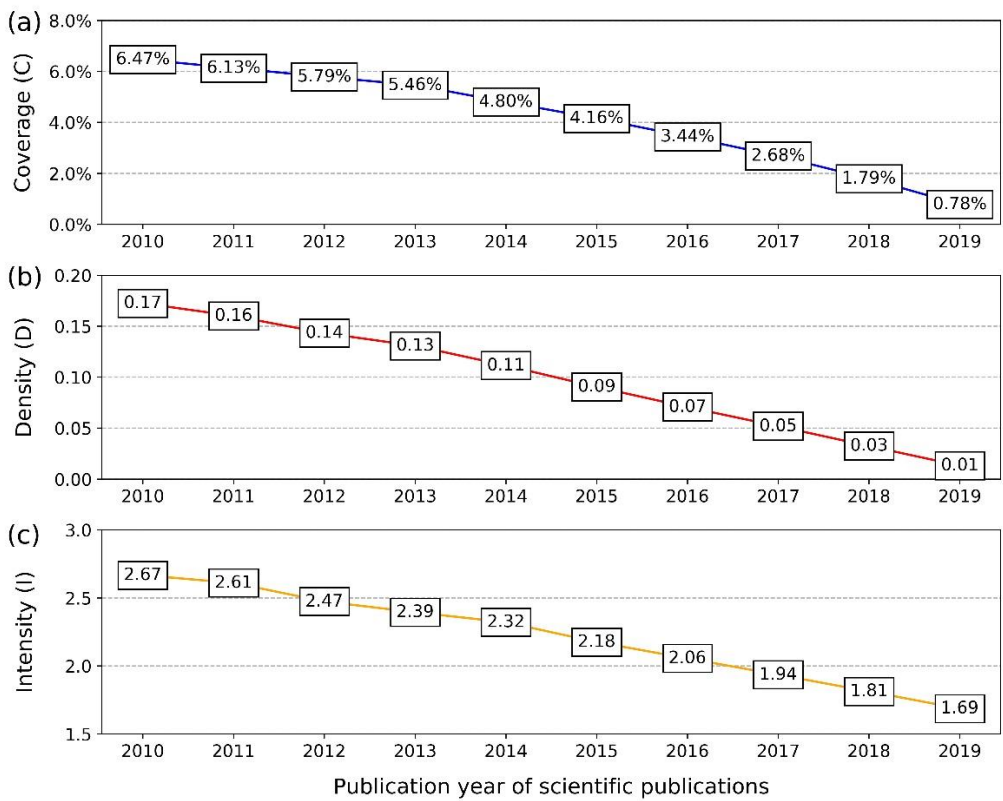

Figure 2. The presence of policy document citations over the publication years of scientific publications measured by (a) Coverage, (b) Density and (c) Intensity

Figure 3 shows the presence of policy document citations across different document types to study which types of publications attract greater attention from policymakers. We find that policy document citations are distributed disproportionately among document types. Reviews are the most present among policy document citations, with over 8% of them being cited by policy documents at least once (C) and are cited 0.18 times on average (D). Articles, otherwise the most predominant document type among scientific publications, rank second in terms of Coverage (4.38%) and Density (0.10), while they received the most policy document citations regarding Intensity (2.37). In addition, editorial materials and news items also exhibit relatively higher probabilities of appearing in policy documents. Except for the above document types, other types of publications are arguably much less visible in the policy landscape. Like that, policy document citations show a pattern similar to scholarly citations of scientific publications, which also concentrate more on reviews and articles over other document types (Haustein et al., 2015; Sigogneau, 2000).

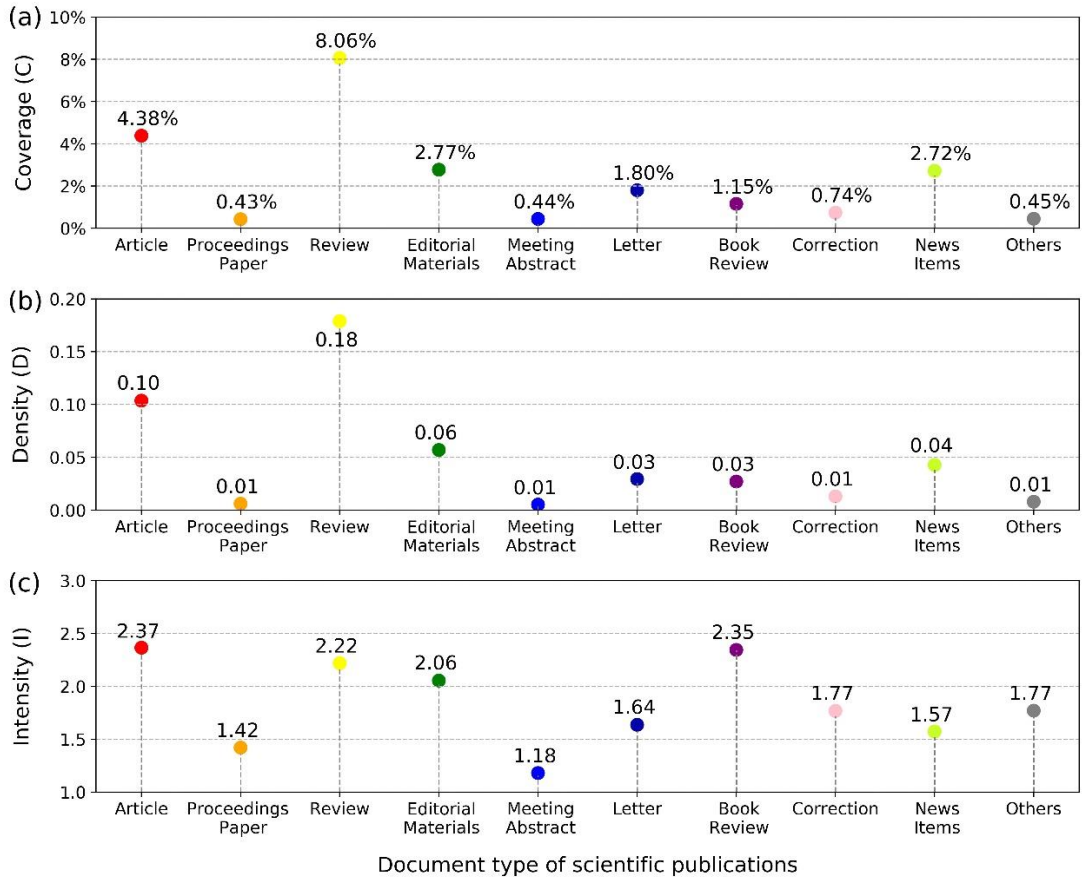

Figure 3. The presence of policy document citations across document types measured by (a) Coverage, (b) Density and (c) Intensity

Figure 4 illustrates the presence of policy document citations across the five main subject fields of science. It shows that SSH publications appear to have the highest probability of being cited by the Overton-indexed policy documents, with about 12.34% of SSH publications cited at least once (C) and 0.45 times on average (D). For those SSH publications with at least one policy document citation, the average number of citations reaches as high as 3.64 (I). In contrast, the presence of policy document citations is also relatively high in the fields of LES (C = 5.96%, D = 0.13, I = 2.11) and BHS (C = 5.72%, D = 0.11, I = 1.89), but much lower in the fields of MCS (C = 0.93%, D = 0.02, I = 1.81) and PSE (C = 0.62%, D = 0.01, I = 1.44). These findings reflect a tendency towards research outputs in the fields of SSH, LES and BHS to attract a greater deal of attention from the policymakers tracked by Overton as compared to the other fields. This pattern was also found for many other altmetric events, such as Twitter mentions, Facebook mentions, and News mentions, etc. (Costas et al., 2015; Fang et al., 2020; Haustein et al., 2015).

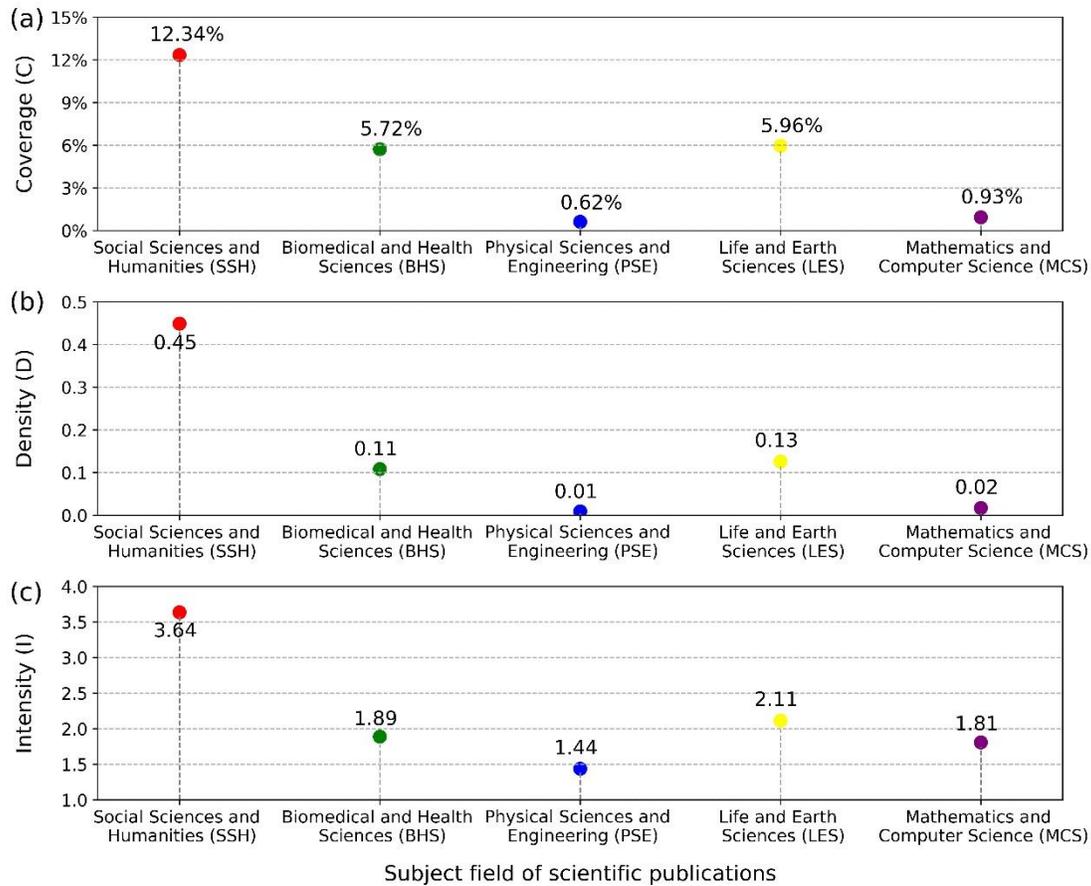

Figure 4. The presence of policy document citations across subject fields measured by (a) Coverage, (b) Density and (c) Intensity

Overlay visualizations created with VOSviewer help to further visualize how policy document citations are distributed across micro-level fields. Figure 5 shows the presence of policy document citations amongst micro-level fields within each subject field of science. Figures 5(b)-(d) are the overlay visualizations of the base map presented in Figure 5(a), in which micro-level fields are scored based upon Coverage, Density and Intensity, respectively. The warmer the color of the node, the higher the value of the corresponding indicator achieved by the micro-level field. It is apparent that the micro-level fields with relatively higher presence of policy document citations, regardless of the indicator, concentrate in the fields of SSH, BHS and LES, thereby confirming that these three subject fields outperform the others in generating policy impact. It should be noted that in a specific subject field, micro-level fields can differ greatly with respect to the presence of policy document citations. This suggests that within a given subject field the Overton-indexed policy documents pay more attention to some research topics than to others.

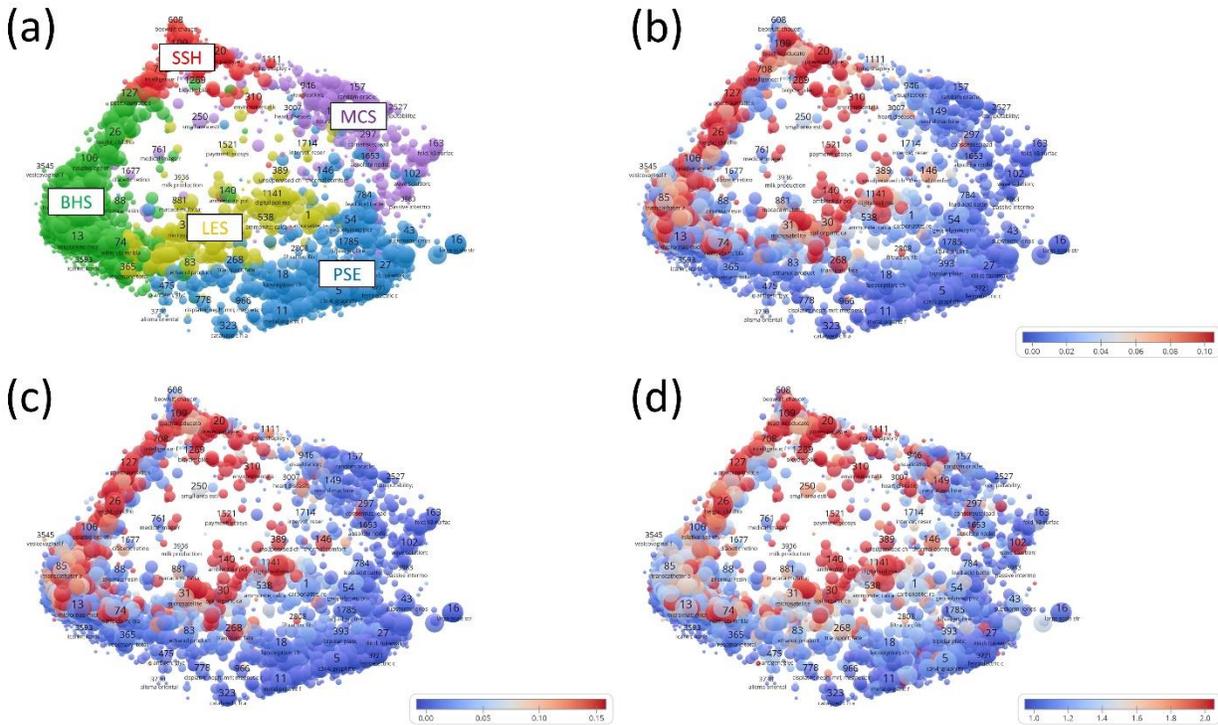

Figure 5. (a) The layout (base map) of the five main subject fields of science; and the overlay visualization scored by (b) Coverage, (c) Density and (c) Intensity, respectively

**Discussion and Conclusion**

This study presents an exploratory large-scale analysis of the presence of policy document citations recorded by Overton for a set of over 18 million WoS publications. Overall, there are about 3.9% of scientific publications having at least one policy document citation. All selected publications have been cited for around 0.09 times on average, and those publications with at least one policy document citation received about 2.32 citations on average. In the dataset at large, the presence of policy document citations is quite sparse, such that the impact on policy cannot be measured properly for the overwhelming majority of publications. In other words, only a limited share of research outcomes was cited by policymakers and thus, can be said to be supportive to policy-making (as captured by policy documents). Given this scarcity of policy document citations across scientific publications, other approaches based on more indirect forms of citation relations – e.g., the ABC approach suggested by Noyons (2019) in which topics are characterized as being relevant for policy, instead of just the cited publications – may become relevant for future policy impact analyses based on the Overton database.

It was also found that the distribution of policy document citations is uneven. First of all, policy document citations are biased towards publications published for a long time, resulting from a citation delay for newly published publications (as in scholarly citations). Given that the accumulation speed of policy document citations is relatively low (Fang & Costas, 2020), it is important to establish an appropriate time window while measuring the policy impact of research outputs. Another similarity between policy document citations and scholarly citations is the preference for articles and reviews. The probability of being cited in policy documents is the highest for reviews, followed by articles, editorial materials and news items. Other document types have very few policy document citations to be studied. This suggests that for policy document citations, articles and reviews are the more relevant document types to be considered when studying scientific impact on policy.

From the perspective of subject field distribution, policy document citations and scholarly citations show different distribution patterns. As already known from previous studies, scholarly citing behavior is more widespread in the fields of natural sciences and medical and health sciences and less prevalent in social sciences and humanities (Fang et al., 2020; Haustein et al., 2015; Marx & Bornmann, 2014), but it seems that for policymakers tracked by Overton there is a tendency to cite more scientific knowledge from the fields of social sciences and humanities, biomedical and health sciences, and life and earth sciences, confirming the same tendency towards society, health, and environment also observed for many other social media metrics (e.g., Twitter mentions, blogs citations, news media mentions) (Fang et al., 2020). This seems to indicate that these research fields are more closely related to policy in the Overton database.

Finally, there are several limitations to be acknowledged to this study. First, the matching process between cited scientific publications and citing policy documents is highly dependent on the existence and identification of DOIs, which essentially excludes all scientific publications without a DOI. Second, we highlighted that policymakers in the Overton database paid unequal attention to research topics within each subject field, yet we didn't answer the question about which specific individual research topics have been of particular interest for policymakers, an aspect that we plan to explore in future research.

**Acknowledgement**

Zhichao Fang is financially supported by the China Scholarship Council (201706060201). Rodrigo Costas is partially funded by the South African DST-NRF Centre of Excellence in Scientometrics and Science, Technology and Innovation Policy (SciSTIP). The authors thank Overton for providing the data for research purposes.

**References**

Adie, E. (2020a). *How far back does the database go?* . http://help.overton.io/en/articles/3831926-how-far-back-does-the-database-go

Adie, E. (2020b). *What is Overton?* http://help.overton.io/en/articles/3822563-what-is-overton

Bornmann, L., Haunschild, R., & Marx, W. (2016). Policy documents as sources for measuring societal impact: how often is climate change research mentioned in policy-related documents? *Scientometrics*, *109*(3), 1477–1495. https://doi.org/10.1007/s11192-016-2115-y

Costas, R., Zahedi, Z., & Wouters, P. (2015). Do "altmetrics" correlate with citations? Extensive comparison of altmetric indicators with citations from a multidisciplinary perspective. *Journal of the Association for Information Science and Technology*, *66*(10), 2003–2019. https://doi.org/10.1002/asi.23309

Fang, Z., & Costas, R. (2020). Studying the accumulation velocity of altmetric data tracked by Altmetric.com. *Scientometrics*, *123*(2), 1077–1101. https://doi.org/10.1007/s11192-020-03405-9

Fang, Z., Costas, R., Tian, W., Wang, X., & Wouters, P. (2020). An extensive analysis of the presence of altmetric data for Web of Science publications across subject fields and research topics.

*Scientometrics*, *124*(3), 2519–2549. https://doi.org/10.1007/s11192-020-03564-9

Haunschild, R., & Bornmann, L. (2017). How many scientific papers are mentioned in policy-related documents? An empirical investigation using Web of Science and Altmetric data. *Scientometrics*, *110*(3), 1209–1216. https://doi.org/10.1007/s11192-016-2237-2

Haustein, S., Costas, R., & Larivière, V. (2015). Characterizing social media metrics of scholarly papers: The effect of document properties and collaboration patterns. *PLoS ONE*, *10*(3), e0120495. https://doi.org/10.1371/journal.pone.0120495

Marx, W., & Bornmann, L. (2014). On the causes of subject-specific citation rates in Web of Science. *Scientometrics*, *102*(2), 1823–1827. https://doi.org/10.1007/s11192-014-1499-9

Noyons, E. (2019). Measuring Societal Impact Is as Complex as ABC. *Journal of Data and Information Science*, *4*(3), 6–21. https://doi.org/10.2478/jdis-2019-0012

Sigogneau, A. (2000). An analysis of document types published in journals related to physics: Proceeding papers recorded in the Science Citation Index database. *Scientometrics*, *47*(3), 589–604. https://doi.org/10.1023/A:1005628218890

Waltman, L., & Van Eck, N. J. (2012). A new methodology for constructing a publication-level classification system of science. *Journal of the American Society for Information Science and Technology*, *63*(12), 2378–2392. https://doi.org/10.1002/asi.22748

Wilson, E. (2020). *What sources does Overton track?* http://help.overton.io/en/articles/3822743-what-sources-does-overton-track

Yu, H., Cao, X., Xiao, T., & Yang, Z. (2020). How accurate are policy document mentions? A first look at the role of altmetrics database. *Scientometrics*, 1–24. https://doi.org/10.1007/s11192-020-03558-7

**Appendix**

Table A1. Distribution of publications and policy document citations in each publication year

| Publication year | Number of publications | Number of publications with at least one policy document citation | Number of policy document citations |
| --- | --- | --- | --- |
| 2010 | 1,420,741 | 91,867 | 245,044 |
| 2011 | 1,518,524 | 93,112 | 243,079 |
| 2012 | 1,623,047 | 93,945 | 232,447 |
| 2013 | 1,710,757 | 93,410 | 223,691 |
| 2014 | 1,800,244 | 86,437 | 200,632 |
| 2015 | 1,877,431 | 78,057 | 169,811 |
| 2016 | 1,988,712 | 68,372 | 140,560 |
| 2017 | 2,046,162 | 54,830 | 106,119 |
| 2018 | 2,126,650 | 38,075 | 69,005 |
| 2019 | 2,230,872 | 17,465 | 29,451 |

Table A2. Distribution of publications and policy document citations in each document type

| Document type | Number of publications | Number of publications with at least one policy document citation | Number of policy document citations |
| --- | --- | --- | --- |
| Article | 13,723,251 | 600,957 | 1,421,998 |
| Proceedings Paper | 1,226,112 | 5,238 | 7,440 |
| Review | 869,945 | 70,133 | 155,642 |
| Editorial Materials | 814,445 | 22,556 | 46,349 |
| Meeting Abstract | 775,569 | 3,410 | 4,022 |
| Letter | 367,561 | 6,608 | 10,812 |
| Book Review | 320,172 | 3,693 | 8,661 |
| Correction | 139,552 | 1,028 | 1,817 |
| News Item | 64,767 | 1,760 | 2,767 |
| Others | 41,766 | 187 | 331 |

Table A3. Distribution of publications and policy document citations in each subject field

| Subject field | Abbr. | Number of publications | Number of publications with at least one policy document citation | Number of policy document citations |
|---|---|---|---|---|
| Social Sciences and Humanities | SSH | 1,351,517 | 166,799 | 606,676 |
| Biomedical and Health Sciences | BHS | 5,917,073 | 338,643 | 639,531 |
| Physical Sciences and Engineering | PSE | 4,741,820 | 29,184 | 41,957 |
| Life and Earth Sciences | LES | 2,202,063 | 131,170 | 277,275 |
| Mathematics and Computer Science | MCS | 1,713,979 | 15,940 | 28,816 |